\documentclass[aps,prb,twocolumn,amsmath,amsfonts,superscriptaddress]{revtex4}
\usepackage{graphicx}
\usepackage{dcolumn}
\usepackage{bm}
\usepackage{boxedminipage}
\usepackage{amsmath}
\usepackage{graphicx}
\usepackage{amsfonts}
\usepackage{feynmf}
\usepackage{mathrsfs}
\usepackage{listings}
\usepackage{algorithm}
\usepackage{algorithmic}
\usepackage{color}

\begin{document}


\title{Disorder-induced exceptional and hybrid point rings in Weyl/Dirac semimetals}


\author{Taiki Matsushita}
\affiliation{Department of Materials Engineering Science, Osaka University, Toyonaka, Osaka 560-8531, Japan}
\author{Yuki Nagai}
\affiliation{CCSE Japan Atomic Energy Agency, 178-4-4, Wakashiba, Kashiwa, Chiba, 277-0871, Japan}
\affiliation{Mathematical Science Team, RIKEN Center for Advanced Intelligence Project (AIP),
1-4-1 Nihonbashi, Chuo-ku, Tokyo 103-0027, Japan}
\author{Satoshi Fujimoto}
\affiliation{Department of Materials Engineering Science, Osaka University, Toyonaka, Osaka 560-8531, Japan}


\date{\today}

\begin{abstract}
Non-Hermiticity in Weyl Hamiltonian leads to the realization of Weyl exceptional rings and flat bands inside the Weyl exceptional rings. Recently, the platform of non-Hermitian physics is extended to many-body or disordered systems where quasiparticles possess finite lifetime. Here, we clarify that the deviation from unitarity limit in a disordered Weyl semimetal leads to the generation of Weyl exceptional rings regardless of the detail of the scattering potential. In the case of topological Dirac semimetals, hybrid point rings and flat bands without vorticity of complex-energy eigenvalues are realized. This scenario is applicable to any Weyl or Dirac semimetals. These effects are detectable by using photoemission or quasiparticle interference experiments.
\end{abstract}


\maketitle

\section{Introduction}
Recently, topology in non-Hermitian systems attracts much attention because of  its  significant effects on electronic structures \cite{shen2018topological,gong2018topological,kawabata2018symmetry,lee2016anomalous,kawabata2019non,hatano1996localization}, and also non-equilibrium dynamics, particularly focusing on $\mathcal{PT}$-symmetric systems\cite{kawabata2018parity}.
Non-Hermiticity of Hamiltonian implies the quantum system has energy gain or loss. 
Therefore, open quantum systems or non-equilibrium systems are the platform of the non-Hermitian quantum physics and have been intensively studied \cite{rotter2009non,daley2014quantum,hatano2019exceptional}. 
Recently, it is found that many-body or disordered systems are also the platform of non-Hermitian physics due to its finite lifetime of excitations \cite{kozii2017non,shen2018topological,papaj2019nodal,zyuzin2018flat,moors2019disorder,yoshida2018non,michishita2019relation,yoshida2019symmetry,kimura2019chiral}.
For many-body or disordered systems, non-Hermiticity of the Hamiltonian appears via the dulled spectrum function and the novel  quasiparticle dispersion. 
Especially, the realization of the exceptional points are widely discussed \cite{kozii2017non,kawabata2019non,yoshida2019symmetry,okugawa2019topological,budich2019symmetry}.  
The band structure of the non-Hermitian Hamiltonian possesses exceptional points,
where the eigenvectors as well as the eigenenergies are degenerate, and thus, they cannot span the full Hilbert space \cite{shen2018topological}. 
This remarkable feature of the non-Hermite energy band is experimentally observable via angular-resolved photoemiession spectroscopy (ARPES) or quasiparticle interference experiment (QPI) measurements. 

A theoretical study of a Weyl fermion system with dissipation, which can be realized in cold atomic systems or photonic systems, showed that spin-dependent gain or loss in this system leads to the formation of a Weyl exceptional ring (WER) and the generation of a flat band inside the WER \cite{xu2017weyl,cerjan2019experimental,mcclarty2019non}. 
For solid state systems, a WER and the flat band inside the WER due to impurity potentials in Weyl semimetals (WSMs) were considered\cite{zyuzin2018flat}. In this previous study, it was shown that the tilt of the Weyl cones in the disordered WSMs leads to the formation of WERs. This result is applicable to disordered WSMs with tilted Weyl cones. 
However, Weyl semimetal materials with large tilts of Weyl cones are rather limited.
Therefore, for the experimental verification of WERs in Weyl semimetals, it is desirable to explore for more general mechanisms of WERs. 

In this paper, we propose the general prescription for the generation of the disorder-induced WERs. 
In our scheme, WERs appear regardless of the detail of scatterers as long as the impurity scatterings are not in the unitarity limit. For example, magnetic impurities or the tilt of Weyl cones are not necessary for its realization.
The condition for the generation of WERs holds ubiquitously, and they are realizable for almost all of WSMs except for very special cases.
Because of the generality and simplicity of our proposal, its experimental detection via ARPES or QPI measurements for  the well-established WSM materials, such as TaAs or Co$_3$Sn$_2$S$_2$, is feasible \cite{lv2015experimental,wang2018large}. 
We, furthermore, investigate a fate of WERs and the flat bands inside WERs at Weyl semimetal-insulator transition induced by tuning symmetry-breaking fields, and propose that the two WERs merge into a hybrid exceptional ring, at the transition point where a topological Dirac semimetal (TDSM) realizes, leading to a flat band without vorticity of the phase of complex-energy eigenvalues inside the hybrid point ring.

This paper is organized as follow. In Sec.~\ref{NH hamiltonian}, we introduce a model for a disordered Weyl semimetal. Here, we define the energy-dependent quasiparticle Hamiltonian and derive the energy-independent effective quasiparticle Hamiltonian, which is generally non-Hermitian. In Sec.~\ref{disorder induced WER}, we discuss disorder-induced WERs by using the self-consistent $T$-matrix approximation. In this section, we verify that as long as the system is deviated from the unitarity limit, WERs generally appear. In addition, we clarify that {\it intra-valley multiple scatterings} play a crucial role for the realization of WERs. In Sec~\ref{Topo-phase transition}, we discuss a fate of WER at topological phase transition point. Here, we show that a hybrid point ring and a flat band without vorticity of the phase of complex-energy eigenvalues are realized just at the transition point.
In Sec~\ref{hybrid point ring TDSM}, we discuss the realization of the hybrid point ring in generic TDSM materials.

\section{Non-Hermitian Quasiparticle Hamiltonian}
\label{NH hamiltonian}
In this section, we define the energy-dependent quasiparticle Hamiltonian, which can be non-Hermitian. By extracting low energy parts in the vicinity of  the poles of the retarded Green's function, we can obtain a general procedure for constructing the effective energy-independent quasiparticle Hamiltonian, which can also be non-Hermitian.

We first introduce the energy-dependent quasiparticle Hamiltonian $\tilde{H}(\epsilon,{\bm p})$: 
\begin{align}
    \tilde{H}(\epsilon,{\bm p}) &= H_0({\bm p}) + \Sigma(\epsilon), \label{eq:nonh}
\end{align}
where $H_0({\bm p}$) is the single-particle Hamiltonian and $\Sigma(\epsilon)$ is the self-energy\cite{shen2018quantum}. With the use of this quasiparticle Hamiltonian, the retarded Green's function can be defined as $G^{\rm R}({\bm p},\epsilon) \equiv [\epsilon -  \tilde{H}(\epsilon,{\bm p})]^{-1}$. 
Because of the imaginary part of the retarded self-energy associated with the finite lifetime of quasiparticles, the energy-dependent Hamiltonian is generally non-Hermitian.

The complex poles of the Green's function is determined by ${\rm det} [\epsilon -  \tilde{H}(\epsilon,{\bm p})]$.
In low energy regions, the self-energy can be written as $ \Sigma(\epsilon) =  \Sigma_0+  \Sigma_1\epsilon +\mathcal{O}(\epsilon^2)$.
In the low energy region, we can introduce the energy-independent quasiparticle Hamiltonian ${\cal H}_{\rm eff}({\bm p})$:
\begin{align}
    {\cal H}_{\rm eff}({\bm p}) &= [1 - \Sigma_1]^{-1} (H_0({\bm p})+  \Sigma_0), \label{eq:heff}
\end{align}
whose eigenvalues ${\cal E}_n({\bm p})$ determine the complex poles of the Green's function because of ${\rm det} [\epsilon -  \tilde{H}({\bm p},\epsilon)] \simeq {\rm det} [\epsilon -{\cal H}_{\rm eff}({\bm p})] =0$.
Note that the quasiparticle Hamiltonian $\tilde{H}(\epsilon,{\bm p})$ and ${\cal H}_{\rm eff}({\bm p})$ can be {\it non-Hermitian} and that its spectrum can be complex.
The non-Hermitian matrix can be non-diagonalizable at certain momenta.
These points are called exceptional points (EPs) in the mathematical physics literature, which are topologically stable.
The topological exceptional points can appear in the quasiparticle spectrum.
In the following sections, we show that these EPs naturally appears in disordered WSMs.

Finally, we comment on the origin of the non-Hermiticity of disordered systems. First of all, we note that the eigenvalue of the quasiparticle Hamiltonian is not the eigenvalues of Hamiltonian, but the poles of the dressed Green's function. Thus, its non-Hermiticity does not mean the existence of the total energy gain and loss in quantum systems. The non-Hermiticity of many body or disordered systems reflects 
finite life-time of quasiparticles, and in the case of multi-band systems such as WSMs, leads to drastic change of the spectral functions of
quasiparticles, 
which are directly observable via ARPES or QPI measurements.

For deeper understanding of non-Hermiticity of disordered system, we consider 
more rigorous approaches, 
e.g. the exact-diagonalization for sufficiently large systems. In a large real-space system, the eigenvalues of the Hamiltonian are real. 
However, we focus on the quasiparticle spectrum obtained from the poles of the Green's function.
More precisely, we can calculate the real-space Green's function $G({\bm x},{\bm x}')$ with impurities and $G_0({\bm x},{\bm x}')=G({\bm x}-{\bm x}')\delta({\bm x}-{\bm x}')$ without impurities. Then, with the use of the Fourier transformation, we obtain the Fourior transformed Green's function  $G({\bm p},{\bm p}') = {\cal FT}[G({\bm x},{\bm x}')]$ and  $G_0({\bm p},{\bm p}') = {\cal FT}[G_0({\bm x},{\bm x}')]$ where ${\cal FT}$ is the Fourier transformation. The self-energy in momentum space is defined by $\Sigma ({\bm p},{\bm p}') =  [G_0^{-1}({\bm p},{\bm p}') ] -[G^{-1}({\bm p},{\bm p}') ]$. If we consider the self-energy with ${\bm p}={\bm p}'$, we can obtain the poles of the Green's function $G({\bm p})  \equiv G({\bm p},{\bm p}) $, which is measured by ARPES experiments.
Since the inverse of the Fourier transformed Green's function $ {\cal FT}[G(({\bm x},{\bm x}') )]$ can have the imaginary part, the self-energy can have the imaginary part. In terms of physical understanding, this means that the one-particle Green's function is smeared with the impurities even in a sufficiently large system. This smearing effect arises from the self-energy. Therefore, even without gain/loss in the entire system, there is a complex spectrum defined by the poles of the dressed Green's function, since the energy gain/loss for quasiparticles exists because of scattering processes.
We also would like to point out that Ref. [12] considered large systems in real space and the authors compared the results between the numerical and analytical calculations. This also means that we can use the complex spectrum of the non-Hermitian quasiparticle Hamiltonian in both numerical and analytical calculations.

\section{Weyl exceptional ring and topological flat band in disordered WSM}
\label{disorder induced WER}
\subsection{Weyl exceptional ring : effective quasiparticle Hamiltonian }
Here, we discuss WERs and disorder-induced tilted Weyl cones by constructing energy-independent effective Hamiltonian. We consider a minimal two band lattice model of WSM. The minimal model of WSM is given by
\begin{align}
    \label{hamWSM}
    H_0({\bm p})= {\bm R}({\bm p}) \cdot {\bm \sigma},
\end{align}
where ${\bm \sigma} = (\sigma_x,\sigma_y,\sigma_z)^T$ is a vector of the Pauli matrices and ${\bm R}({\bm p}) = (v \sin p_x, v \sin p_y, \gamma (\cos p_z -m))^T$. This simple model  of a WSM describes low energy physics of two adjacent WPs. For example, if we regard that the Pauli matrices is in the orbital space, this two-band model can describe the two WPs near $L$-point in Weyl ferromagnet, Co$_3$Sn$_2$S$_2$ \cite{ozawa2019two}.

The model Eq.~(\ref{hamWSM}) exhibits the transition between the WSM phase and the band insulator phase, depending on the parameter $m$. 
In the case that $|m| < 1$, WPs appear in ${\bm p}=(0,0,\pm \cos^{-1} m)$. At $|m|=1$, the positive and negative chiral fermions overlap in the Brillouin zone (BZ).  
For $|m|>1$, the band gap opens, resulting in the band insulator.

 We, now, consider effects of disorder potentials in this system, which are incorporated into the self-energy.
As will be shown later, the self-energy is expressed as $\Sigma(\epsilon) = \Sigma_0(\epsilon) \sigma_0 + \Sigma_z(\epsilon) \sigma_z$. To obtain the energy-independent quasiparticle Hamiltonian, we expand it in the energy $\epsilon$ as
\begin{eqnarray}
 \Sigma_0(\epsilon) &\equiv& \Sigma_0^0 + \Sigma_0^1\epsilon+\mathcal{O}(\epsilon^2),\\
 \Sigma_z(\epsilon) &\equiv& \Sigma_z^0 + \Sigma_z^1 \epsilon+\mathcal{O}(\epsilon^2).
\end{eqnarray}
As shown in Eq.~(\ref{eq:heff}), the quasiparticle Hamiltonian in low-energy region ${\cal H}_{\rm eff}({\bm p})$ is written as 
\begin{align}
     {\cal H}_{\rm eff}({\bm p})= {\bm R}'({\bm p}) \cdot {\bm \sigma} + ab(p_z) \sigma_0,
\end{align}
where
\begin{eqnarray}
{\bm R}'({\bm p}) &=& (aR_x'({\bm p}),aR_y'({\bm p}),aR_z'({\bm p})),
\end{eqnarray}
\begin{eqnarray}
R_x'(p_x,p_y)&=&(1-\Sigma_0^1)R_x(p_x)+i\Sigma_z^1R_y(p_y),\\
R_y'(p_x,p_y)&=&(1-\Sigma_0^1)R_y(p_y)-i\Sigma_z^1R_x(p_x),\\
R_z'(p_z)&=&(1-\Sigma_0^1)(R_z(p_z)+\Sigma_z^0)+\Sigma_z^1\Sigma_0^0,
\end{eqnarray}
with the renormalizatiuon coefficient $a \equiv 1/ \left( (1-\Sigma_0^1)^2-(\Sigma_z^1)^2) \right)$ and $b(p_z) \equiv  (1-\Sigma_0^1)\Sigma_0^0+\Sigma_z^1(R_z(p_z)+\Sigma_z^0)$. The $b(p_z)$-term leads to tilted of Weyl cone \cite{papaj2019nodal}.

The eigenvalue of ${\cal H}_{\rm eff}$ is expressed as 
\begin{align}
    {\cal E}_\pm({\bm p}) &= a\left(\pm \sqrt{ R_x'(p_x,p_y)^2 +  R_y'(p_x,p_y)^2 + R_z'(p_z)^2} +b(p_z)\right). \label{eq:ep}
\end{align}
The non-Hermitian Hamiltonian is not diagonalizable when the first term in Eq.~(\ref{eq:ep}) is zero since there is only one eigenvector.
The exceptional points form WER on ${\bm p}_d$, where the equations 
${\rm Re} \: [R_z'(p_{dz})] = 0$ and 
$R_x'(p_{xd},p_{yd})^2+R_y'(p_{xd},p_{yd})^2 = {\rm Im} \: [ R_z'(p_{dz})]^2$ are satisfied.
The above discussion indicates that the imaginary part of $\Sigma_z(\epsilon)$ has an important role for the appearance of  the WER.

\subsection{Disorder-induced Weyl exceptional ring}
Here, we clarify that WERs appear regardless of the detail of the scattering potentials, as long as there are {\it intra-valley multiple scattering processes}.
As shown below, neither the orbital dependence of scattering potentials nor magnetic impurities is necessary, in contrast to two-dimensional Dirac systems, where these specific features of scattering potentials are necessary for the realization of exceptional rings\cite{kozii2017non}. 

We consider randomly distributed impurities and assume the scattering potential $V({\bm x})=V_0(\sigma_0+\beta \sigma_z)\sum_a \delta({\bm x}-{\bm x}_a) $. Here, ${\bm x}_a$ expresses a site of scatterers and $\beta$ is difference of scattering potential between two degrees of freedom (orbital or spin). 
Especially, if Pauli matrices is in spin space, the finite $\beta$ means magnetic impurities.
The self-energy $\Sigma(\epsilon)$ derived from the self-consistent $T$-matrix approximation is given as
\begin{eqnarray}
\label{selfenergy}
\Sigma(\epsilon)&=&n_{\rm imp}T(\epsilon)=\Sigma_0(\epsilon)\sigma_0+\Sigma_z(\epsilon)\sigma_z,\\
\label{T}
T(\epsilon)&=&T_0(\epsilon)\sigma_0+T_z(\epsilon)\sigma_z\nonumber\\
&=&V_0\left(\sigma_0+ \beta \sigma_z \right) \nonumber \\
&+&V_0\left(\sigma_0+ \beta \sigma_z \right)\int_{\rm BZ} \frac{d^3 p}{(2\pi )^3}G(\epsilon,{\bm p})T(\epsilon),
\end{eqnarray}
where $T(\epsilon)$ and $n_{\rm imp}$ are the $T$-matrix and the concentration of scatterers, respectively. Here, we can decompose the integrated Green's function as $\int_{\rm BZ} \frac{d^3 p}{(2\pi )^3}G(\epsilon,{\bm p})\equiv \overline{G}_0(\epsilon)+\overline{G}_z(\epsilon)\sigma_z$. The off-diagonal components of the integrated Green's function are canceled since the off-diagonal components of Hamiltonian is odd functions of momentum. The $\Sigma_0$ term broadens the spectrum and does not qualitatively change the dispersion of quasiparticles. On the other hand, $\Sigma_z$ term drastically changes the dispersion relation of quasiparticles as discussed in the previous section.
The expression of integrated Green function is 
\begin{eqnarray}
\label{integrated G0}
\overline{G}_0&=&\int_{\rm BZ} \frac{d^3 p}{(2\pi )^3}\frac{\epsilon-\Sigma_0}{\tilde{E}_+(\epsilon,{\bm p})-\tilde{E}_-(\epsilon,{\bm p})}\nonumber\\&\times&\left(\frac{1}{\epsilon-\tilde{E}_+(\epsilon,{\bm p})}-\frac{1}{\epsilon-\tilde{E}_-(\epsilon,{\bm p})}\right),\\
\label{integrated Gz}
\overline{G}_z&=&\int_{\rm BZ} \frac{d^3 p}{(2\pi )^3}\frac{ \gamma \cos p_z-m+\Sigma_z}{\tilde{E}_+(\epsilon,{\bm p})-\tilde{E}_-(\epsilon,{\bm p})}\nonumber\\&\times&\left(\frac{1}{\epsilon-\tilde{E}_+(\epsilon,{\bm p})}-\frac{1}{\epsilon-\tilde{E}_-(\epsilon,{\bm p})}\right).
\end{eqnarray}
where $\tilde{E}_\pm(\epsilon,{\bm p})$ is the complex eigen-energy of the energy-dependent quasiparticle Hamiltonian $\tilde{H}(\epsilon,{\bm p})$ defined in Eq.~(\ref{eq:nonh}), which is given by
\begin{eqnarray}
&&\tilde{E}_\pm(\epsilon,{\bm p}) =\nonumber\\ &\pm& \sqrt{(\gamma(\cos p_z-m)+\Sigma_z(\epsilon))^2+v^2(\sin^2 p_x+\sin^2 p_y)}+\Sigma_0.\nonumber\\
\end{eqnarray}
For $\epsilon={\rm Re}\Sigma_0(\epsilon),\; p_z=\pm \cos^{-1} (m-{\rm Re}\Sigma_z(\epsilon))/\gamma,\; \sin^2 p_x+\sin^2 p_y =|{\rm Im} \Sigma_z(\epsilon)/v|$, the energy-dependent quasiparticle Hamiltonian is not diagonalizable. These exceptional points form a WER. In the surface area enclosed by this WER
$\sin^2 p_x+\sin^2 p_y <|{\rm Im} \Sigma_z(\epsilon)/v|$, the real part of complex-energy eigenvalue vanishes and flat band characterized by vorticity of the phase of the complex-energy eigenvalues is realized. 
\begin{figure}[t]
\includegraphics[width=8cm]{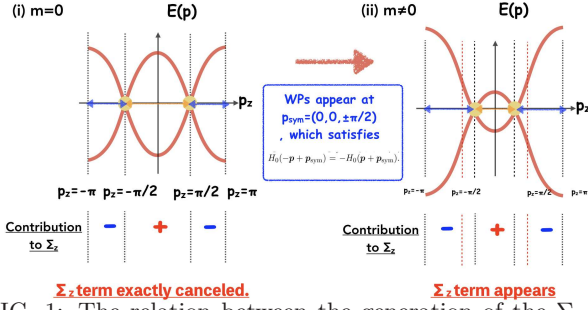}
\vspace{-0.6cm}
\caption{The relation between the generation of  the $\Sigma_z$ and the parameter $m$.}
\label{fig01}
\end{figure}
\begin{figure}[b]
\includegraphics[width=7cm]{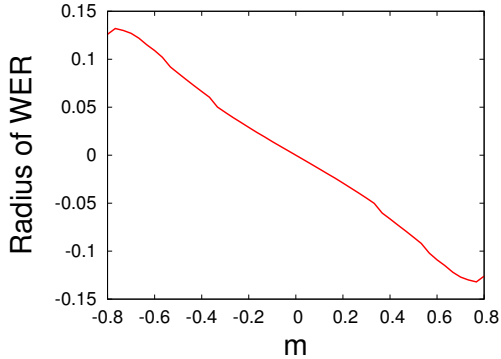}
\caption{The $m$-dependence of the radius of a WER. The parameters are set as, $\gamma=1,\; v=0.6,\; n_{\rm imp}=0.05,\; V_0=3,\; \beta=0$. In this paper, we set the $p$-mesh size $35 \times 35 \times 35$. }
\label{fig0}
\end{figure}
The self-consistent equations (\ref{T}) can be transformed as
\begin{eqnarray}
\label{Sigma_0}
\Sigma_0(\epsilon)=n_{\rm imp}V_0\frac{1-(1-\beta^2)V_0\overline{G}_0}{\left( 1-V_0\left(\overline{G}_0+\beta \overline{G}_z\right) \right)^2-V_0^2(\overline{G}_z+\beta \overline{G}_0)^2},\nonumber\\\\
\label{Sigma_z}
\Sigma_z(\epsilon)=n_{\rm imp}V_0\frac{\beta+(1-\beta^2)V_0\overline{G}_z}{\left( 1-V_0\left(\overline{G}_0+\beta \overline{G}_z\right) \right)^2-V_0^2(\overline{G}_z+\beta \overline{G}_0)^2}.\nonumber\\
\end{eqnarray}
In the similar manner to the two dimensional Dirac system, the orbital dependence of scattering potential or magnetic impurities leads to the generation of exceptional points \cite{kozii2017non,papaj2019nodal}. Interestingly, even when scattering potential does not have orbital dependence or magnetic character, i.e. $\beta=0$, 
$\Sigma_z(\epsilon)$ naturally appears in the case that $m\neq 0$. To see this more clearly, we consider isotropic scattering potentials $\beta=0$. In this case, the self-consistent equation (\ref{Sigma_z}) becomes
\begin{eqnarray}
\label{Sigma_z_nonb}
\Sigma_z(\epsilon)&=&n_{\rm imp}V_0\frac{V_0\overline{G}_z}{\left( 1-V_0\overline{G}_0 \right)^2-V_0^2\overline{G}_z^2}.
\end{eqnarray}
Eq. (\ref{integrated Gz}) shows $\overline{G}_z$ becomes finite in the case that the parameter $m$ is non-zero. 
Therefore, we can conclude that WERs can be realized in the case of $m\neq0$ even when $\beta=0$.

Here, we clarify the physical meaning of the parameter $m$, which is related to one of the conditions for the generation of WERs.
The parameter $m$ lifts two cosine bands on $p_x=p_y=0$ and determines the position of WPs. 
The finite $m$ means that the position of WPs deviates from the symmetric points of BZ : ${\bm p}_{\rm sym}=(0,0,\pm \pi/2)$, which is defined by 
\begin{eqnarray}
H_0(-{\bm p}+{\bm p}_{\rm sym})=-H_0({\bm p}+{\bm p}_{\rm sym}).
\end{eqnarray}
As shown in FIG.~\ref{fig01}, when WPs are located on the symmetric points of the BZ, the contributions to the integrated Green's function $\overline{G}_z$ from the region $p_z>|p_{z{\rm sym}}|$ and the region $p_z<|p_{z{\rm sym}}|$ are exactly canceled.
FIG.~\ref{fig0} shows the $m$-dependence of the radius of WER , which indicates that 
 the radius increases as the function of the distance between the WPs and the symmetric points.
Therefore, the deviation of the position of WPs from the symmetric points of the BZ is a key ingredient for the realization of large WERs. 
 We stress that this condition for large WERs are satisfied for many of candidate materials of WSMs, and $m=0$ is rarely realized in real systems. 

\subsection{Origin of WER and disappearance of WER at unitarity limit}
In this subsection, we verify that the physical origins of WERs found in the previous subsection are {\it intra-valley multiple scattering processes}.
We also discuss disorder-induced tilted Weyl cones and their asymmetric density of states (DOS). 
Moreover, we show that WERs disappear in the unitarity limit. 

\begin{figure}[t]
\includegraphics[width=9cm]{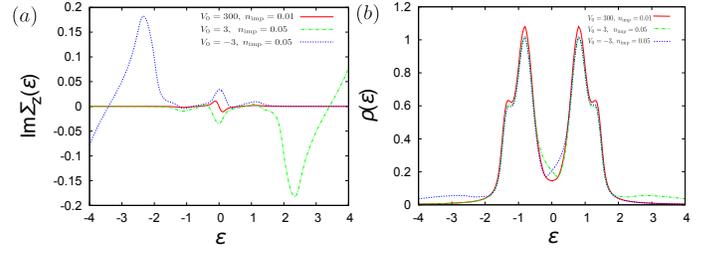}
\caption{Figure (a) and  (b) are the $\epsilon$ dependence of ${\rm Im}\Sigma_z(\epsilon)$ and the density of states $\rho(\epsilon)$ respectively. The red (solid line), green (dashed line) and blue (dotted line) curves correspond to $(V_0/\gamma,n_{\rm imp})=(300,0.01),\;(3,0.05),\;(-3,0.05)$. The other parameters are set as, $\gamma=1,\; v=0.6,\; m=0.3,\beta=0$.} \label{fig1}
\end{figure}
\begin{figure}[b]
\includegraphics[width=9cm]{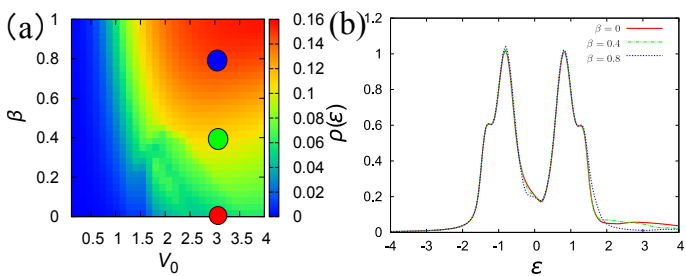}
\caption{(a) Radius of a WER, $|{\rm Im}\Sigma_z(\epsilon)/v|$, at $\epsilon={\rm Re}\Sigma_0(\epsilon)$. The parameters are the same as those used in FIG.1. (b) DOS at the three points indicated by colors, red (solid line), green (dashed line), and blue (dotted line), in FIG.3(a) : $(V_0/\gamma,\beta)=(3,0),\;(3,0.4),\;(3,0.8)$. } 
\label{fig2}
\end{figure}
\begin{figure}[b]
\includegraphics[width=9cm]{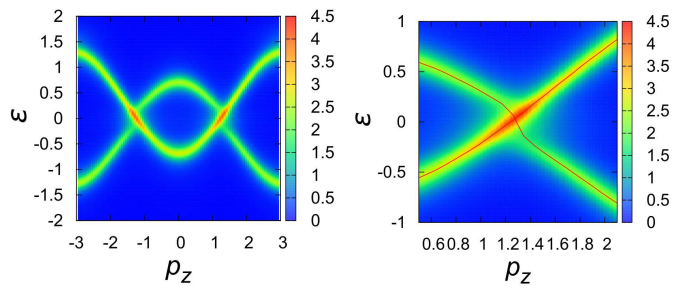}
\caption{The left figure is the spectral function at $p_x=p_y=0$. The right one is the enlarged figure around the Weyl points. The red line indicates the dispersion relation of quasiparticles. It shows that the disorder-induced tilted of Weyl cone appears.}
\label{fig3}
\end{figure}
\begin{figure*}[t]
 \begin{center}
 \includegraphics[width=16cm]{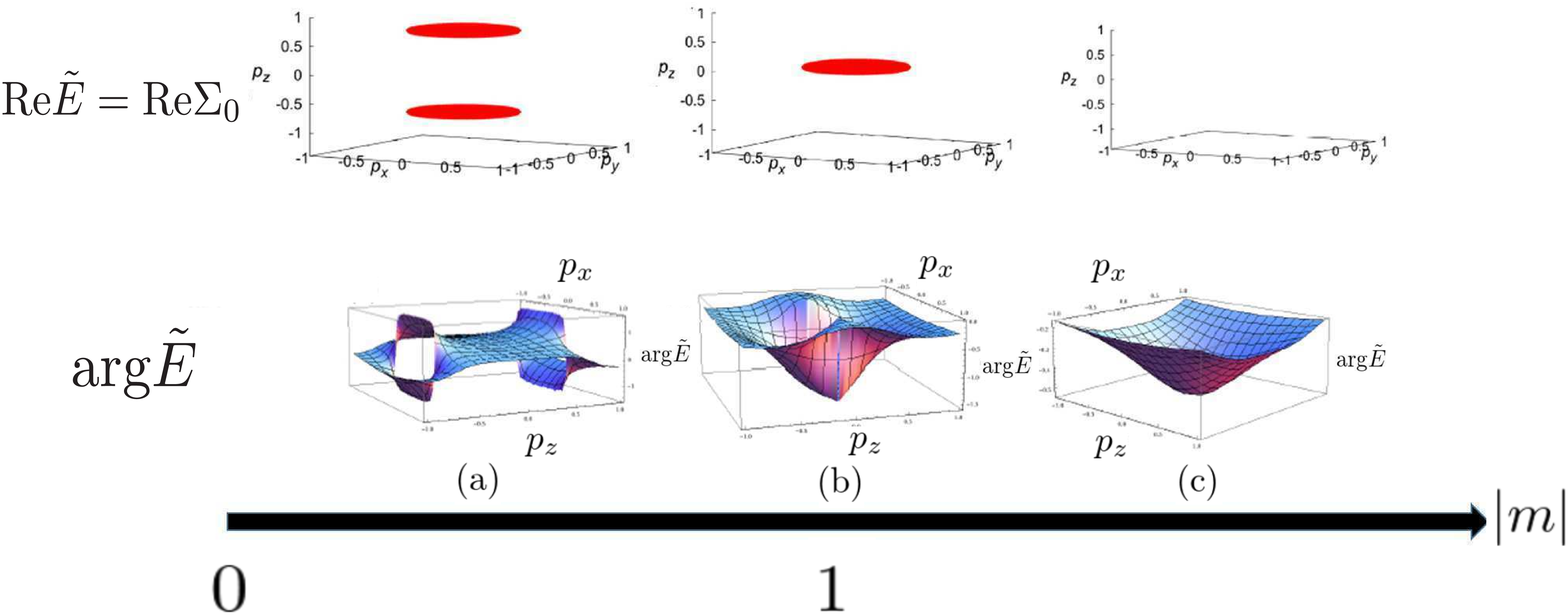}
 \caption{Zero energy points ${\rm Re}\tilde{E}({\bm p})=0$ and the phase of the energy eigenvalues : (a) WSM phase ($|m|<1$), (b) phase boundary ($|m|=1$), (c) insulator phase ($|m|>1$).}
 \label{fig5}
 \end{center}
\end{figure*}

FIG.\ref{fig1} (a) and (b) show that the $\epsilon$-dependence of ${\rm Im}\Sigma_z(\epsilon)$ and the density of states (DOS)  $\rho(\epsilon)$, respectively. It is shown that disorder makes the DOS asymmetric. This asymmetric behavior depends on the sign of scattering potentials, which implies that the asymmetry is due to the formation of impurity bands. The red curves (solid line) in FIG.\ref{fig1} (a) is ${\rm Im}\Sigma_z(\epsilon)$ in the unitarity limit : $V_0\to \infty ,\; n_{\rm imp}\to 0$. In this limit, the symmetry of DOS $\rho(-\epsilon)=\rho(\epsilon)$ is recovered. In the unitarity limit, the self-consistent equations for (\ref{Sigma_0}) and (\ref{Sigma_z}) become
\begin{eqnarray}
\label{Sigma_0unitarity}
\Sigma_0(\epsilon)&=&-n_{\rm imp}V_0\frac{\overline{G}_0(\epsilon)}{\overline{G}_0(\epsilon)^2-\overline{G}_z(\epsilon)^2},\\
\label{Sigma_zunitarity}
\Sigma_z(\epsilon)&=&n_{\rm imp}V_0\frac{\overline{G}_z(\epsilon)}{\overline{G}_0(\epsilon)^2-\overline{G}_z(\epsilon)^2}.
\end{eqnarray}
In the unitarity limit, the self-consistent equations no longer depend on $\beta$. Here, we assume the $\epsilon$-dependence of the integrated Green's function as : $\overline{G}_0(-\epsilon)=-\overline{G}_0^*(\epsilon),\;\overline{G}_z(-\epsilon)=\overline{G}_z^*(\epsilon)$. From Eqs.~(\ref{Sigma_0unitarity}) and (\ref{Sigma_zunitarity}), we can easily find $\Sigma_0(-\epsilon)=-\Sigma_0^*(\epsilon),\;\Sigma_z(-\epsilon)=\Sigma_z^*(\epsilon)$. These conditions and Eqs. (\ref{Sigma_0unitarity}) and (\ref{Sigma_zunitarity}) are self-consistent because $\tilde{E}_-(-\epsilon,\bm{p})=-\tilde{E}_+^*(\epsilon,\bm{p})$ is satisfied. Therefore, in the unitarity limit, ${\rm Im}\Sigma_z(\epsilon)=0$ at $\epsilon={\rm Re}\Sigma(\epsilon)=0$, which implies that WERs do not appear\cite{com1}.

As discussed in Appendix, within the self-consistent Born approximation, WERs do not appear for $\beta=0$\cite{suppliment}. The disappearance of WER in the unitarity limit or in the Born approximation implies that the origins of the generation of WERs are multiple scattering processes with moderate strength of impurity potentials. 
In addition, as discussed in Appendix, we found that intra-valley scatterings lead to the generation of WER. Therefore, we conclude {\it intra-valley multiple scatterings} in WSM generally leads to the realization of WERs\cite{suppliment}. Combining the results obtained in the previous section, we can conclude that  
the deviation from the unitarity limit and the deviation of WPs from the symmetric points of the BZ are crucial conditions for realization of WERs. 
This is our main result. 
These conditions are general and do not depend on the details of scattering potentials. 
Neither magnetic impurities nor the orbital dependence of scattering potentials are necessary. For most of real candidate materials of WSMs, WPs are not located on symmetric points of the BZ. Therefore, for almost all Weyl material, WERs can be realized by generic disorder potentials, and the experimental detection of them are feasible.

In FIG.~\ref{fig2} (a), the radius of WER ($|{\rm Im}\Sigma_z(\epsilon)/v|$ at $\epsilon={\rm Re}\Sigma_0(\epsilon)$) is shown. 
The results in FIG.\ref{fig2} (a) show that the radius of the WER is increased by $\beta$ in the manner similar to the two-dimensional Dirac system \cite{papaj2019nodal}. 

Usually, the generation of a flat band is accompanied with the singular behavior of the DOS\cite{marchenko2018extremely}. 
In FIG.~\ref{fig2} (b), the DOS for three different values of $\beta $ are shown.
In FIG.\ref{fig2} (b), we can not see the signature of the realization of flat bands. As discussed in Appendix, the growth of a flat band inside the WER leads to the increase of the DOS. However, it is relatively small \cite{marchenko2018extremely}.
This is because that  the formation of the flat band inside the WER can be regarded as split of WPs. Therefore, the increment of DOS due to the generation of flat band inside WER is negligibly small in as shown in FIG.\ref{fig2} (b). 
The asymmetry of the DOS near $\epsilon \simeq 0$ is originated from energy dependence of the self energy and disorder-induced tilted Weyl cones. Actually, as shown in FIG.~\ref{fig3}, disorder potentials induce the tilts of Weyl cones, which breaks the symmetry of the DOS: $\rho(-\epsilon)=\rho(\epsilon)$. 

Here, we discuss the difference between our study and previous ones. Ref. [13] discusses WSMs with tilted Weyl cones and shows that the tilt of Weyl cones leads to the realization of WERs. This scenario is applicable to multilayer system or Type II WSM. On the other hand, our proposal is applicable to more general disordered WSM materials. 

We comment on the relation between our two band model and real candidate materials. If we regard that the Pauli-matrices is in orbital space, our two band model can be seen as the effective Hamiltonian of Weyl ferromagnet Co$_3$Sn$_2$S$_2$\cite{xu2017weyl,ozawa2019two}. Actually, our two band model can describe the low energy physics of this material governed by WPs near $L$-points. The simple band structure derived from half-metalic character and low carrier density of this material can be considered as that Co$_3$Sn$_2$S$_2$ is good platform for demonstration of Weyl physics in condensed matter\cite{liu2018giant}. Therefore, the experimental verification of WERs via ARPES or QPI measurements is feasible.

To close this section,  we discuss the relation between the generation of WERs and a disorder-driven semimetal to  diffusive metal transition. In WSMs, the semimetalic featrure of the DOS ($\rho(\epsilon)\propto \epsilon^2$)  is robust against weak disorder~\cite{fradkin1986critical}. Within the perturbative approach, a certain strength of disorder potentials leads to the phase transition to a diffusive metal phase~\cite{kobayashi2014density,bera2016dirty,shapourian2016phase}. In the semimetalic region, WERs can not be realized because the complex selfenergy can not be introduced due to the causality. As the asymmetric DOS shows (FIG.~\ref{fig3}), the state obtained in our calculation is in a diffusive metallic phase. Thus WERs are realized. Recently, it is verified that there is a non-perturbative contribution to the DOS which renders the semimetal state unstable~\cite{pixley2016rare,pixley2017single}. We can expect that WERs can be realizable even in this region~\cite{com2}.

\section{fate of WER and flat band at topological phase transition point}
\label{Topo-phase transition}
In this section, we discussed a fate of WERs at a topological phase transition point. We introduce the self-energy $\Sigma(\epsilon) \simeq \Sigma_0 \sigma_0+\Sigma_z \sigma_z$ by hand in the two band model which is discussed in the previous section. 

In FIG. \ref{fig5}, we show disorder-induced flat bands inside WER and the phase of complex-energy eigenvalues for the WSM phase $|m|<1$,  the phase boundary $|m|=1$ and the band insulator phase $|m|>1$. We can see that flat bands without vorticity of the phase of complex energy eigenvalues is realized at the topological phase transition point. At the phase transition point, a vortex and an anti-vortex of the phase of the complex-energy eigenvalues are overlapped and the vorticities are canceled. However, at the topological phase transition point, non-Hermitian Hamiltonian is still non-diagonalizable at the edge of the flat band. 
The set of $\bm{p}$-points at the edge of the flat band constitutes a hybrid point ring.
In contrast to EPs which are accompanied by a quantized vortex of the phase of complex-energy eigenvalues, hybrid points are not accompanied by a vortex. Hybrid points are also topological defects of the energy spectrum of non-Hermitian systems. 
As shown in FIG.~\ref{fig5}, the hybrid point ring is realized at the topological phase transition point. 

\section{hybrid point ring in topological Dirac semimetal}
\label{hybrid point ring TDSM}
In previous section, we demonstrate that merging of two WERs leads to the realization of a hybrid point ring. 
In this section, we verifies its realization in TDSM systems.

We consider Cd$_3$As$_2$ which is a typical TDSM system with a simple band structure and topologically protected Dirac points (DPs)\cite{borisenko2014experimental,kobayashi2015topological}. Since these topologically protected DPs can not be gapped out as long as the four-fold rotational symmetry is retained, this material can be considered as an ideal platform for the realization of hybrid point ring and flat band without vorticity of the phase of the complex-energy eigenvalues. The effective Hamiltonian of this material is the same as the sum of two systems of the lattice model considered in Sec.~\ref{disorder induced WER}.  
\begin{figure}[t]
\includegraphics[width=8cm]{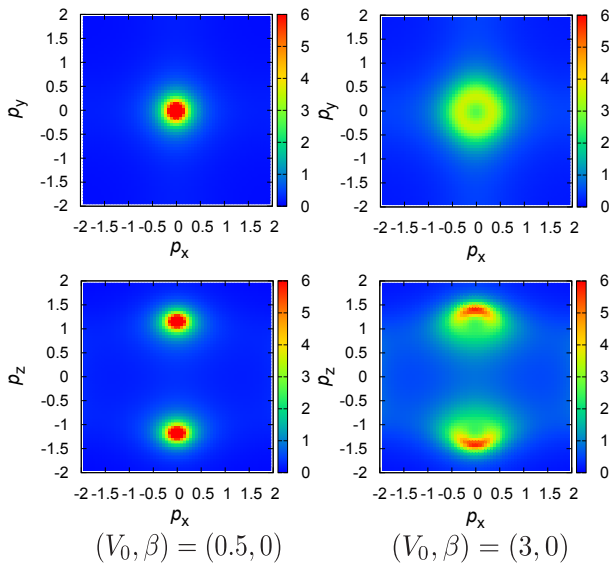}
\caption{Spectral functions of a TDSM. The left and right figures show the spectral functions at $V_0/|t_0|=0.5,\;3$ respectively. The upper panel shows the spectral functions at $\epsilon={\rm Re}\Sigma_0(\epsilon),\;p_z=\cos^{-1}\left( -(t_0+2t_2+{\rm Re}\Sigma_z(\epsilon))/t_1 \right)$. The lower panel shows the spectral functions at $\epsilon={\rm Re}\Sigma_0(\epsilon),\; p_y=0$. The parameters are set to $n_{\rm imp}=0.05,\; \beta=0 ,\;V_0=3,\;V_0=1,\; t_0=-1,\;t_1=1,\;t_2=0.3,\;\Lambda=0.7$.}
\label{fig6}
\end{figure}
The effective Hamiltonian is given by
\begin{eqnarray}
\label{hamiltonian TDSM}
H_0^{{\rm TDSM}}({\bm p})=
\begin{pmatrix} 
h({\bm p})&&0\\
0&&-h({\bm p})
\end{pmatrix}
=h({\bm p})s_{z},
\end{eqnarray}
where $h({\bm p})=m_p\sigma_z+\Lambda(\sigma_x \sin p_xa+\sigma_y \sin p_ya),\;m_k=t_0+t_1 \cos p_za +t_2 (\cos p_xa+\cos p_ya)$ \cite{wang2013three}. $\sigma_i\;(i=x,y,z)$ and $s_i\;(i=x,y,z)$ are Pauli matrices in orbital and spin space, respectively. $t_1$ and $t_2$ are hopping integrals.
$\Lambda$ describes hybridization of $s$ and $p$ orbital due to anti-symmetric spin-orbit interaction. The effective Hamiltonian can be regarded as the sum of two WSM models with opposite chirality. The sum of positive and negative chiral Fermions leads to DPs, which appear at ${\bm p} = (0,0,\pm \cos^{-1} (-(t_0+2t_2)/t_1))$.

To verify the realization of the hybrid point ring in the TDSM, we introduce disorder potentials in this model by using the self-consistent $T$-matrix approximation. We consider the scattering potential $V({\bm x})=V_0(\sigma_0+\beta \sigma_z)\sum_a \delta({\bm x}-{\bm x}_a) $. Here, ${\bm x}_a$ expresses a site of scatterers and $\beta$ is the difference between scattering potentials of $s$ and $p$ orbitals. In FIG.~\ref{fig6}, the spectral function of the TDSM is shown. The flat band appears even when $\beta=0$. Asymmetry of scattering potentials in orbital space increases the radius of the hybrid point ring in a manner similar to the case of two-band Weyl model discussed in the previous sections.

\section{Conclusion}
In this paper, we have investigated disorder-induced WERs, clarifying the general scheme for their realization. 
In our scheme, WERs are realized regardless of the details of scattering potentials, and also  the tilt of the Weyl cone is not necessary. 
The position of the WPs  and multiple scattering  processes play important roles for  the generation of the WERs.
The conditions for the realization WERs are summarized as follows: (i) the positions of WPs deviate from the symmetric points of the BZ, (ii) The scattering potentials deviate from the unitarity limit. 
We note that the condition (i) is satisfied for almost all candidate materials of WSMs.
Therefore, we can conclude that WERs are realizable almost all WSMs and experimentally detectable. 
In particular, our two-band model is relevant to the Weyl ferromagnet, Co$_3$Sn$_2$S$_2$. 

In addition, we have verified that the generation of the flat band inside the WER leads to the small increase of the DOS. 
Moreover, the energy dependence of the self-energy leads to the disorder-induced tilted Weyl cone and asymmetry of the DOS. These features can be experimentally verified by STM measurements. 

Finally, a fate of WER at a topological phase transition point was discussed. 
We have found that, at the transition point where the TDSM state is realized, merging of two WERs with a vortex and an anti-vortex of complex energy eigen-values leads to the realization of the hybrid point ring and topological flat band without vorticity. 
We have demonstrated the realization of the hybrid point ring and the flat band in the case of Cd$_3$As$_2$. 

Our theoretical proposal can directly be detected via ARPES or QPI measurement.
These measurement are useful for detection of the anisotropic broadening of quasiparticle spectrum which is related evolving exceptional or hybrid rings.  We believe that it become a large step for non-Hermitian physics in disordered systems.

\section{Acknowledgments}
TM thanks T. Mizushima, K. Nomura, A. Ozawa, Y. Ominato and K. Kobayashi for faithful discussion. TM acknowledges illuminating discussions with T. Yoshida and J. H. Pixley,. 
The calculations were partially performed by the supercomputing system SGI ICE X at the Japan Atomic Energy Agency. Especially, J. H. Pixley inform us of Refs. [36] and [37], which are useful for the deeper understanding our results.
TM was supported by a JSPS Fellowship for Young Scientists.
YN was partially supported by the ''Topological Materials Science'' (No. 18H04228) KAKENHI on Innovative Areas from JSPS of Japan.
SF was supported by the Grant-in-Aids for Scientific
Research from JSPS of Japan (Grants No. 17K05517), and KAKENHI on Innovative Areas ``Topological Materials Science'' (No.~JP15H05852) and "J-Physics" (No.~JP18H04318). 

\appendix

\section{Complex energy-eigen value of $\epsilon$-independent quasiparticle Hamiltonian}
Here, we demonstrate that energy dependence of self-energy leads to tilted Weyl cones. Considering low energy regions, we expand the self-energy as,
\begin{eqnarray}
\Sigma(\epsilon)&=&\Sigma_0(\epsilon)\sigma_0+\Sigma_z(\epsilon)\sigma_z
\nonumber\\&=&(\Sigma_0^0+\Sigma_0^1\epsilon)\sigma_0+(\Sigma_z^0+\Sigma_z^1\epsilon)\sigma_z+\mathcal{O}(\epsilon^2),
\end{eqnarray}
where the upper subscript shows the order of frequency. There is no off-diagonal component of the self-energy because the off-diagonal components of Hamiltonian are odd functions of momentum. 

The complex energy eigen-value equation of the 2-band Weyl model is given as,
\widetext
\begin{eqnarray}
{\rm det}\left(\epsilon-H_0({\bm p})-\Sigma(\epsilon)\right)&=&
\begin{vmatrix}
(1-\Sigma^1_0-\Sigma^1_z)\epsilon-h(p_z)-\Sigma_0^0-\Sigma_z^0&&-v\left(\sin p_x-i \sin p_y \right) \\
-v\left(\sin p_x+i \sin p_y\right)&&(1-\Sigma^1_0+\Sigma^1_z)\epsilon+h(p_z)-\Sigma_0^0+\Sigma_z^0
\end{vmatrix}=0,
\end{eqnarray}
where we define $h(p_z)=\gamma(\cos p_z-m)$. The dispersion relation is derived from this equation. 
The complex energy eigenvalue is given as,
\begin{eqnarray}
\label{tiltedcone}
E_\pm({\bm p})&=&a\left(\pm \sqrt{b(p_z)^2+a^{-1}\left[(h(p_z)+\Sigma_z^0)^2+v^2\left(\sin^2 p_x+\sin^2 p_y\right)\right]}+b(p_z) \right),
\end{eqnarray}
\endwidetext
with $a \equiv 1/ \left( (1-\Sigma_0^1)^2-(\Sigma_z^1)^2) \right)$ and $b(p_z) \equiv (1-\Sigma_0^1)\Sigma_0^0+\Sigma_z^1(R_z(p_z)+\Sigma_z^0)$. 
$\Sigma_0^1$ can be incorporated as 
the renormalization of the Fermi velocity and band width. 
On the other hand, $\Sigma^1_z$ changes qualitatively the dispersion relation of quasiparticles. We also found that $\Sigma_z(\epsilon)$ leads to tilting of the Weyl cone.

\section{Disappearance of Weyl exceptional ring within Born approximation}\label{WERSCBA}
Here, we verify that Weyl exceptional ring does not appear within the Born approximation. We consider the two-band model discussed in the main text. 
\begin{eqnarray}
\label{ham}
H_0({\bm p})=\gamma(\cos p_z-m)\sigma_z+v(\sin p_x\sigma_x+\sin p_y\sigma_y),
\end{eqnarray}
We assume the orbital dependent impurity potential $V({\bm x})=V_0(\sigma_0+\beta \sigma_z)\sum_a \delta({\bm x}-{\bm x}_a) $. The self-energy within the self-consistent Born approximation is given by
\begin{eqnarray}
\label{SCBA}
\Sigma_{\rm SCBA}(\epsilon)&=&n_{\rm imp}V_0(\sigma_0+\beta \sigma_z)\nonumber\\
&+&n_{\rm imp}V_0^2(\sigma_0+\beta \sigma_z)\overline{G}(\epsilon)(\sigma_0+\beta \sigma_z),
\end{eqnarray}
where $\overline{G}(\epsilon)\equiv \frac{d^3p}{(2\pi)^3}G(\epsilon,{\bm p})$ is the integrated Green's function. 
Because the off-diagonal components of the Hamiltonian are odd functions of momentum, the integrated Green's function is diagonal. Therefore, it is written as $\overline{G}(\epsilon)=\overline{G}_0\sigma_0+\overline{G}_z\sigma_z$. 
This means that the self-energy is also given as,
\begin{eqnarray}
\Sigma_{\rm SCBA}(\epsilon)=\Sigma_{{\rm SCBA}\;0}(\epsilon)\sigma_0+\Sigma_{{\rm SCBA}\;z}(\epsilon)\sigma_z.
\end{eqnarray}
The self-consistent equation (\ref{SCBA}) gives
\begin{eqnarray}
\label{self-SCBA}
\Sigma_{{\rm SCBA}\;0}(\epsilon)&=&n_{\rm imp}V_0^2\left[ (1+\beta^2)\overline{G}_0(\epsilon)+2\beta\overline{G}_z(\epsilon) \right],\nonumber\\\\
\Sigma_{{\rm SCBA}\;z}(\epsilon)&=&n_{\rm imp}V_0^2\left[ 2\beta\overline{G}_0(\epsilon)+(1+\beta^2)\overline{G}_z(\epsilon) \right].\nonumber\\
\end{eqnarray}
The integrated Green's function is given by
\begin{eqnarray}
\label{integrated G0 SCBA}
\overline{G}_0&=&\int_{\rm BZ} \frac{d^3 p}{(2\pi )^3}\frac{\epsilon-\Sigma_{{\rm SCBA} \;0}}{\tilde{E}_+(\epsilon,{\bm p})-\tilde{E}_-(\epsilon,{\bm p})}\nonumber\\&\times&\left(\frac{1}{\epsilon-\tilde{E}_+(\epsilon,{\bm p})}-\frac{1}{\epsilon-\tilde{E}_-(\epsilon,{\bm p})}\right),\\
\label{integrated Gz SCBA}
\overline{G}_z&=&\int_{\rm BZ} \frac{d^3 p}{(2\pi )^3}\frac{ \gamma \cos p_z-m+\Sigma_{{\rm SCBA}\;z}}{\tilde{E}_+(\epsilon,{\bm p})-\tilde{E}_-(\epsilon,{\bm p})}\nonumber\\&\times&\left(\frac{1}{\epsilon-\tilde{E}_+(\epsilon,{\bm p})}-\frac{1}{\epsilon-\tilde{E}_-(\epsilon,{\bm p})}\right).
\end{eqnarray}
where $\tilde{E}_\pm(\epsilon,{\bm p})$ is the complex energy eigenvalue of non-Hermitian Hamiltonian $\tilde{H}({\bm p})\equiv H_0({\bm p})+\Sigma_{\rm SCBA}(\epsilon)$, which is given as,
\widetext
\begin{eqnarray}
\label{comp-energy}
\tilde{E}_\pm(\epsilon,{\bm p}) =\pm \sqrt{(\gamma(\cos p_z-m)+\Sigma_{{\rm SCBA}\;z}(\epsilon))^2+v^2(\sin^2 p_x+\sin^2 p_y)}+\Sigma_{{\rm SCBA}\;0}(\epsilon).
\end{eqnarray}
\endwidetext

In the case of $\beta=0$, $\Sigma_{{\rm SCBA}\;0}(\epsilon)\propto \overline{G}_{0}(\epsilon),\; \Sigma_{{\rm SCBA}\;z}(\epsilon)\propto \overline{G}_{z}(\epsilon)$ are satisfied. Here, we postulate the $\epsilon$-dependence as,
\begin{eqnarray}
\label{epdepg0}
\overline{G}_0(-\epsilon)&=&-\overline{G}_0^*(\epsilon),\\
\label{epdepgz}
\overline{G}_z(-\epsilon)&=&\overline{G}_0^*(\epsilon),\\
\label{epdeps0}
\Sigma_{{\rm SCBA}\;0}(-\epsilon)&=&-\Sigma_{{\rm SCBA}\;0}^*(\epsilon),\\
\label{epdepsz}
\Sigma_{{\rm SCBA}\;z}(-\epsilon)&=&\Sigma_{{\rm SCBA}\;z}^*(\epsilon).
\end{eqnarray}
The series of equations (\ref{self-SCBA})-(\ref{integrated Gz SCBA}) and the energy dependence of the integrated Green's function and the self-energy (\ref{epdepg0})-(\ref{epdepsz}) are self-consistent. We have numerically checked the ansatz (\ref{epdepg0})-(\ref{epdepsz}). The, we end up with,
\begin{eqnarray}
{\rm Re}\Sigma_0(\epsilon=0)=0,\;{\rm Im}\Sigma_z(\epsilon=0)=0.
\end{eqnarray}
This means that Weyl exceptional ring does not appear for $\beta=0$ within the self-consistent Born approximation.

\section{The origin of Weyl exceptional ring}
Here, we discuss that the origin of disorder-induced Weyl exceptional rings and show that intra-valley scatterings lead to the generation of Weyl exceptional rings.
\begin{figure}[b]
 \begin{center}
 \includegraphics[width=8cm]{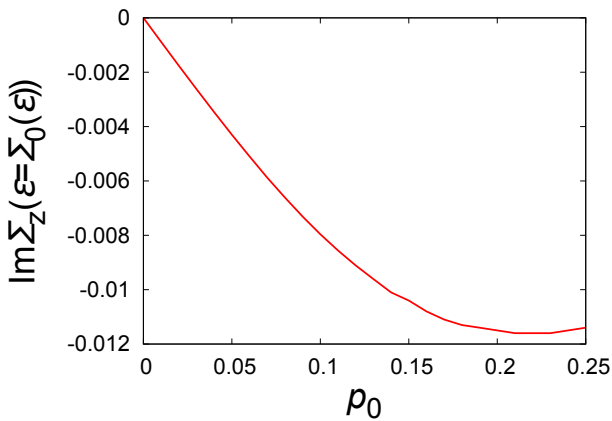}
 \caption{$p_0$-dependence of a radius of Weyl exceptional ring. We set impurity potential as: $n_{\rm imp}=0.03\;,V_{\rm intra}=3,\;V_{\rm inter}=0$.}
 \label{supfig1}
 \end{center}
\end{figure}
\begin{figure}[t]
 \begin{center}
 \includegraphics[width=8cm]{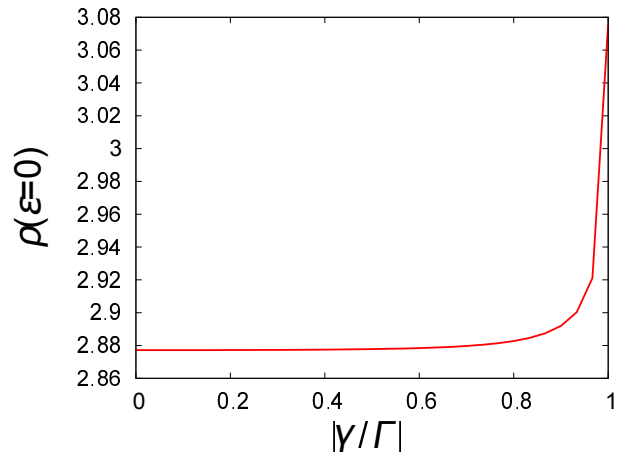}
 \caption{$|\gamma/\Gamma|$-dependence of the DOS $\rho(\epsilon=0)$. we set $\Gamma=0.1$ and momentum cut-off as $p_{\rm cut}=3$.}
 \label{supfig2}
 \end{center}
\end{figure}
To identify the origin of the Weyl exceptional ring, we introduce the valley index $\tau_z={\rm diag}(1,-1)$ and consider a valley-separated model:
\begin{eqnarray}
H_0({\bm p})=\tau_z (p_z-\tau_z p_0) \sigma_z+(p_x\sigma_x+p_y\sigma_y),
\end{eqnarray}
where we choose the position of WPs as ${\bm p}_0=(0,0,\pm p_0)$. 
From now on, we introduce scattering potentials as
\begin{eqnarray}
V({\bm x})=\sum_a\delta({\bm x}-{\bm x}_a)\left( V_{\rm intra}\sigma_0 \tau_0+V_{\rm inter}\sigma_0 \tau_x \right),
\end{eqnarray}
where $\tau_i\; (i=x,y,z)$ are the Pauli matrices of the valley space. Here, $V_{\rm intra}$ and $V_{\rm inter}$ are intra-velley and inter-valley scattering potentials, respectively. 

We apply the self-consistent $T$-matrix approximation. The self-energy and the $T$-matrix are given by,
\begin{eqnarray}
\Sigma(\epsilon)&=&n_{\rm imp}T(\epsilon),
\end{eqnarray}
\begin{eqnarray}
T(\epsilon)&=&\left( V_{\rm intra}\sigma_0 \tau_0+V_{\rm inter}\sigma_0 \tau_x \right)\nonumber\\&+&\left( V_{\rm intra}\sigma_0 \tau_0+V_{\rm inter}\sigma_0 \tau_x \right)\int \frac{d^3p}{(2\pi)^3}G(\epsilon,{\bm p})T(\epsilon).\nonumber\\
\end{eqnarray}
We can easily find that the inter-valley scattering leads to $\Sigma(\epsilon)\propto \tau_x,\;\sigma_z\tau_x$ which opens energy gaps at Weyl points. 

The intra-valley scattering generates Weyl exceptional rings. In FIG.~\ref{supfig1}, we show the the $p_0$-dependence of a radius of the Weyl exceptional ring. It is shown that the separation of Weyl points increases the radius of the Weyl exceptional ring. As we demonstrated above, Weyl exceptional rings do not appear within the self-consistent Born approximation. Therefore, we can conclude that that the separation of Weyl points and intra-valley scatterings are the necessary and sufficient conditions for the realization of Weyl exceptional rings in generic Weyl semimetals.

\section{The increase of the density of states due to the appearance of the Weyl exceptional ring}
Here, we clarify that the DOS is increased by the appearance of WERs. To discuss the relation between the DOS at the energy of the WER and the radius size of it, we consider the simple energy-independent quasiparticle-Hamiltonian,
\begin{eqnarray}
\mathcal{H}_{\rm eff}({\bm p})=-i\Gamma+(p_z-i\gamma)\sigma_z+p_x\sigma_x+p_y\sigma_y,
\end{eqnarray}
where $\sigma_i \;(i=x,y,z)$ are the Pauli matrices, and we assume $\Gamma, \gamma \in \mathbb{R}$. Because of the causality, $\Gamma \geq 0$ and $|\gamma|\leq \Gamma$ are assumed. In this model, a WER appear for $p_x^2+p_y^2 < \gamma^2$ and the flat band is generated inside the WER. Usually, the flat band has the large DOS. Therefore, it is natural to consider the relation between the radius of the WER and the increase of the DOS.

We plot the DOS $\rho(\epsilon=0)$ as a function of $|\gamma/\Gamma|$ in FIG.~\ref{supfig2}. We can see that the DOS is increased by the generation of the WER and the flat band. The increase of the DOS due to the generation of the flat band inside the WER is moderate compared with the case of a usual flat band (for instance, that in the case of bilayer graphene system) \cite{marchenko2018extremely}.

\bibliography{referencenonhermiteWeyl}

\begin{thebibliography}{41}
\bibitem[{com({\natexlab{a}})}]{com1}
\bibinfo{note}{We numerically checked the absence of other solutions.}
\bibitem[{com({\natexlab{b}})}]{com2}
\bibinfo{note}{As discussed in Ref. [33], non-perturbative contributions to the
  DOS have weak frequency dependence. Phenomenologically, it can be described
  by the finite DOS, namely the smearing factor of the bare Green's function.
  We found that the finite DOS due to the smearing fator leads to the
  generation of WERs.}
\end{thebibliography}
\bibliographystyle{apsrev}
\end{document}